\documentclass[aps,prl,twocolumn,showpacs]{revtex4}
\usepackage{graphicx}
\usepackage{amsmath}
\usepackage{amssymb}
\usepackage{amsfonts}
\usepackage{bm}

\begin{document}

\title{Creation and detection of a 
mesoscopic gas in a non-local quantum superposition
}

\author{Christoph Weiss}
\affiliation{Laboratoire Kastler Brossel, \'Ecole Normale
Sup\'erieure, UPMC and CNRS, 24 rue Lhomond, 75231 Paris Cedex 05, France}

\author{Yvan Castin}
\affiliation{Laboratoire Kastler Brossel, \'Ecole Normale
Sup\'erieure, UPMC and CNRS, 24 rue Lhomond, 75231 Paris Cedex 05, France}

\begin{abstract}
We investigate the scattering of 
a quantum matter wave soliton on a barrier in a one dimensional geometry and we
show that it can lead to mesoscopic Schr\"odinger cat states, where the atomic
gas is in a coherent superposition of being in the 
half-space to the left of the barrier 
and being in the half-space to the right of the barrier. 
We propose an interferometric method to reveal the coherent 
nature of this superposition
and we discuss in details the experimental feasibility.
\end{abstract}

\date{\today}

\maketitle

It is now possible to control the strength of the atomic interaction
in a gas, with Feshbach resonances. 
This has allowed the observation of single matter wave bright solitons with
thousands of atoms
\cite{Khaykovich} or a train of solitons \cite{Hulet} 
with $^7$Li atoms trapped in a one-dimensional (1D) geometry.  
These solitons are quantum bound states of a mesoscopic gas, 
which opens up fascinating possibilities:
Apart from testing mean field predictions in
these systems \cite{classical_soliton},
one can address truly quantum problems,
issuing from the quantum nature of the gas center of mass.

In particular, it was recently proposed to use a Bose-Einstein condensate
in interferometric experiments to test the 
existence of decoherence mechanisms
not predicted by usual quantum mechanics 
and that would show up for very massive particles
\cite{Reynaud}. 
Experiments have succeeded in observing interferences with molecules as 
big as fullerenes 
and there is a need for more massive interferometric objects 
\cite{more_massive}.
A soliton with a small number of $100$ $^7$Li atoms 
has the same mass as C$_{60}$,
with appealing new features: 
It does not have internal bound states other than its
ground state, it can be reversibly dissociated in an unbound atomic gas
{\sl via} a Feshbach resonance,
and it allows the exploration of a new regime,
in which the center of mass kinetic energy of the interfering object
is of the same order as the binding energy of its constituents. 

Furthermore, thanks to the extremely
low temperatures accessible in atomic gases, 
down to 0.45nK \cite{Ketterle_record}, and the
weak decoherence present in these systems \cite{Bloch_revivals}, 
one may hope to split the center of mass wavefunction
of the solitonic gas in two wavepackets that would keep their 
mutual coherence over mesoscopic distances,
say a fraction of a millimeter, much larger than the 
size of the soliton. 
The gas would then have simultaneously non-zero probability 
amplitudes of being in two different spatial locations,
thus forming a mesoscopic Schr\"odinger cat in {\sl real} space.
One may then ascertain the presence of a cat state by recombining 
and interfering these
two mesoscopically different quantum states of the gas.
This would constitute a generalization to many atoms of the 
one-ion experiment of \cite{Monroe}.
While mesoscopic Schr\"odinger cat states have been 
reported for radiation fields \cite{Brune}
they have not been reported yet with ultracold atoms, and
atom optics with a quantum soliton is a promising alternative 
to existing ideas for cat production in these systems \cite{cats_bec}.

The dynamics of the center of mass wavepacket during the
scattering of the soliton on a barrier raises
non trivial theoretical issues, since the presence 
of the barrier makes the 1D many-body problem 
non integrable {\sl via} the Bethe Ansatz. 
We thus construct 
an approximate effective low-energy Hamiltonian 
for the center of mass of the gas, and 
we derive a rigorous upper bound on the resulting error.

The starting point is the many-body Hamiltonian in 1D, 
for $N$ bosonic particles of
mass $m$ interacting {\sl via} the usual contact interaction of coupling constant $g$,
in presence of the barrier potential $U(x)\geq 0$:
\begin{equation}
\label{eq:H}
H = \sum_{i=1}^{N} \left[\frac{p_i^2}{2m}+U(x_i)\right]
+g\sum_{i<j} \delta(x_i-x_j).
\end{equation}
This is conveniently rewritten as
$H = P^2/(2M) + H_{\rm in} + V$,
singling out the kinetic energy
of the center of mass ($M=Nm$ is the total mass and $P$ the total momentum
of the gas), the so-called {\sl internal} Hamiltonian $H_{\rm in}$ 
and the sum of the $N$ barrier potentials, $V$. 
Without a barrier
($V\equiv 0$) there is full separability between the center of mass 
and the internal variables, so that 
we split the Hilbert space as a tensorial product of center of mass 
and internal variables.
$H_{\rm in}$ is diagonalized with the Bethe ansatz \cite{Herzog}:
For $N$ fixed, its ground state is its single discrete eigenstate,
the quantum soliton $|\phi\rangle$ of energy $E_0(N)$ 
\cite{McGuire}, separated from a continuum of solitonic fragments
by an energy gap which is minus the chemical potential,
\begin{equation}
|\mu|= E_0(N-1)-E_0(N) = \frac{m g^2 N(N-1)}{8\hbar^2}.
\end{equation}
In presence of a barrier, we consider the scattering state $|\Psi\rangle$
of the soliton with an incoming center of mass wavevector
$K>0$. We restrict to a {\sl low} incoming kinetic energy
to have {\sl elastic} scattering,
\begin{equation}
E-E_0 \equiv \frac{\hbar^2 K^2}{2M} < |\mu|. \label{eq:elastic}
\end{equation}
Far from the barrier, one can then
observe only a non-fragmented soliton, to the right
with the transmission amplitude $t$, to the left with the reflection
amplitude $r$.

\noindent In this elastic regime, an effective hermitian Hamiltonian 
may be defined, introducing the projector
$\mathcal{P} = I_{\rm CoM} \otimes |\phi\rangle \langle \phi|$
acting as the identity on the center of mass and projecting
the internal state on its ground state, so that
\begin{equation}
\mathcal{P} |\Psi\rangle = |\Phi\rangle \otimes |\phi\rangle.
\end{equation}
Far from the barrier,  $\Phi(X)$ is simply the center of mass
wavefunction, $X$ being the center of mass position. 
The so-called
$\mathcal{P}G\mathcal{P}$ formalism, 
where $G$ is the resolvent of the full Hamiltonian
\cite{CCT}, then gives the exact equation 
\begin{equation}
\frac{\hbar^2 K^2}{2M} |\Phi\rangle =
\left[\frac{P^2}{2M} + \bar{V}(X) + \delta V\right] |\Phi\rangle.
\label{eq:exact}
\end{equation}
The first contribution to the effective potential, in the right hand side
of (\ref{eq:exact}), is the convolution of the barrier potential
with the internal density profile of the soliton:
\begin{equation}
\bar{V}(X) = \langle \phi| V|\phi\rangle = \int_{-\infty}^{+\infty} 
dx\  U(X-x) \rho(x|0)
\end{equation}
where $\rho(x|0)$ is the mean density of particles in the soliton
{\sl knowing that}
the center of mass is localized in $X=0$.
It was calculated with the Bethe ansatz 
\cite{Calogero} and is well approximated for $N\gg 1$ by the mean field
density profile
$\rho(x|0) \simeq N/[4\xi \cosh^2(x/2\xi)]$,
where the mean field soliton size is $\xi=\hbar^2/(m|g|N)$.
The second contribution in (\ref{eq:exact})
involves virtual transitions to internal excited states:
\begin{equation}
\delta V = \langle\phi| V \mathcal{Q} 
\frac{\mathcal{Q}}{E\mathcal{Q}-\mathcal{Q}H\mathcal{Q}}
\mathcal{Q} V |\phi\rangle
\end{equation}
where $\mathcal{Q}=I-\mathcal{P}$. We shall neglect this contribution
but not without a justification. From the fact that 
$\mathcal{Q}H\mathcal{Q}\geq E_0+|\mu|$, a consequence of the
positivity of $P^2/2M$ and $V$, and of the energy gap 
of $H_{\rm in}$, we see in the regime (\ref{eq:elastic})
that the operator $-\delta V$ is positive and bounded as
\begin{equation}
-\delta V \leq W(X) \equiv \frac{\langle\phi|V^2|\phi\rangle
-\bar{V}(X)^2}{|\mu|-\hbar^2 K^2/2M}.
\end{equation}
When one neglects $\delta V$ in (\ref{eq:exact}), the exact $\Phi(X)$
is replaced by $\Phi_0(X)$, which involves the same incoming
wave $e^{iKX}$, but with outgoing waves $e^{iK|X|}$ whose transmission
and reflection amplitudes $t_0$ and $r_0$ are only approximate.
We have rigorously bounded the resulting errors.
We discuss here only the experimentally relevant case of an
even barrier $U(x)=U(-x)$. Introducing the ``small parameter", 
$\epsilon \equiv 
M \langle \Phi_0|W(X)|\Phi_0\rangle/(\hbar^2 K |t_0|)$,
we have for $\epsilon<1/2$ the following {\sl theorem}:
\begin{equation}
\label{eq:bound_tr}
|t-t_0| \ \ \mbox{and}\ \ |r-r_0| \leq \frac{|t_0|\epsilon}{1-2\epsilon} .
\end{equation}
It remains to calculate $W(X)$.
We have derived from the Bethe ansatz
the large $N$ asymptotic expression \cite{static}
\begin{eqnarray}
W(X) &\simeq& \frac{2N\xi^4}{|\mu|-\frac{\hbar^2K^2}{2M}}
\int_{-\infty}^{+\infty}\! dx \int_{x}^{+\infty}\! dy
\, U''(X+x\xi)  \nonumber\\
&\times& U''(X+y\xi) \frac{2+y-x}{(e^y+1)(e^{-x}+1)}.
\label{eq:Wasymp}
\end{eqnarray}
In practice, the barrier $U(x)$ is 
produced with a Gaussian laser beam, $U(x)=U_0\exp(-2x^2/b^2)$,
with a waist $b$ much larger 
than the soliton size $\xi$. Then the mean potential $\bar{V}(X)$
is close to $NU(X)$. We shall also assume that the 
incoming kinetic energy $\hbar^2 K^2/2M$ 
is about half the gap $|\mu|\simeq \hbar^2/8m\xi^2$, 
so that
(\ref{eq:elastic}) is satisfied without paying the price of 
very slow soliton velocities. Then $K b\gg 1$
and the scattering is in the semi-classical regime \cite{Berry},
where approximate expressions can be obtained for $t_0$ and $r_0$.
A transmission probability $1/2$ is predicted to be achieved
for an incident wavevector $K_0$ such that
\begin{equation}
\frac{\hbar^2 K_0^2}{2M} = \max_X \bar{V}(X) \simeq N U_0.
\label{eq:htpc}
\end{equation}
In the vicinity of $K=K_0$,
the transmission probability varies sharply from zero
to unity,
\begin{equation}
|t_0|^2 \simeq \frac{1}{1+\exp[\frac{K_0-K}{\delta K}]}
\ \ \mbox{with} \ \
\delta K \simeq \frac{1}{\pi\sqrt{2}b}.
\end{equation}
It remains to estimate the bound (\ref{eq:bound_tr}).
One may take $U''\simeq U''(X)$ in (\ref{eq:Wasymp}),
since $b\gg \xi$, so that
\begin{equation}
W(X) \simeq 
\frac{N\xi^4}{|\mu|-\frac{\hbar^2K^2}{2M}}
\left[U''(X)\right]^2 \left[\frac{2\pi^2}{3} +4\zeta(3)\right].
\end{equation}
In $K=K_0$, for $\epsilon \ll 1$,  a semi-classical calculation 
gives
\begin{equation}
|t-t_0| \lesssim \frac{10(\xi/b)^3}{N^{1/2}}\ln\left(Nb^2/\xi^2\right),
\end{equation}
a quantity checked to be $\ll 1$ in what follows.

We now study the experimental feasibility.
An axial Gaussian laser beam 
confines $N\simeq 100$ atoms of $^7$Li 
in the $y-z$ plane, with a
resulting transverse harmonic oscillator length
$a_\perp=(\hbar/m\omega_\perp)^{1/2}\simeq 0.54 \mu$m, where
$\omega_\perp\simeq 2\pi \times 4.8$KHz is the transverse oscillation frequency.
In this optical wave guide, the interacting gas has a one dimensional
character if $2\xi \gg a_{\perp}$. In order to make cooling of the gas not
too challenging, we take a not too large soliton length
$\xi\simeq 0.9 \mu$m; the resulting 3D scattering length,
$a \simeq -a_\perp^2/(2N\xi) \simeq -1.72$nm is in the interval
of values $(-\infty,-1.5\mbox{nm})$
accessible with the Feshbach resonance \cite{Khaykovich}.
Initially the gas is also harmonically trapped along $x$ with
an oscillation frequency $\omega$. The gas is assumed to be cooled
to the temperature $T=0.45$nK \cite{Ketterle_record}.
This axial trap is so weak that it very weakly affects
the internal solitonic variables, $\hbar \omega < |\mu|/10$,
but it is strong enough that the center of mass of the gas, still
separable in a harmonic trap, has a negligible
probability $\exp(-\hbar\omega/k_B T)<1/10$ to be in an excited state. 
These two constraints impose the weak value
$\omega\simeq 2\pi\times 23.5$Hz. They also imply $|\mu|/k_B T\simeq 25$,
so that the internal variables of the soliton are frozen in their ground state.

At $t=0$, the gas is launched with a total momentum $\hbar K_0$ such
that
\begin{equation}
\frac{\hbar^2 K_0^2}{2M} = \frac{|\mu|}{2} \simeq \frac{\hbar^2}{16m\xi^2}.
\end{equation}
The corresponding velocity is $\hbar K_0/M \simeq 0.37$mm/s.
Simultaneously the axial trap is switched off, to free
the center of mass of the gas, with an initial wavepacket
\begin{equation}
\Phi(X) \propto e^{iK_0 X} e^{-(X-X_0)^2 (\Delta K)^2}.
\label{eq:init}
\end{equation}
With a sudden opening of the axial trap \cite{cree_exc}, 
$\hbar^2 (\Delta K)^2/2M
=\hbar \omega/4$ and $\Delta K/K_0 \simeq 0.22 \ll 1$: 
This wavepacket is quasi-monochromatic.
Smaller values of $\Delta K$ may be obtained
by a more clever opening procedure of the trap, within
times $\sim 1/\omega$ \cite{Bouchoule}.
The wavepacket is then scattered on a broad Gaussian barrier centered
in $x=0$ (here $X_0<0$),
a {\sl beam-splitter},
created by a laser beam of waist $b=5\xi\gg \xi$ and of intensity
adjusted to satisfy the half-transmission probability condition
(\ref{eq:htpc}). In any realistic case,
$\Delta K$ remains much larger than $\delta K$,
so the wavepacket experiences a mere filtering in Fourier space, the components
with $K>K_0$ being transmitted and the ones with $K<K_0$ being reflected \cite{fluctu}.
As a consequence, the wavepacket also splits in real space
in a transmitted part and a reflected part,
that nicely separate since their mean velocity
exceeds their spreading velocity: A mesoscopic Schr\"odinger cat is born.

How to prove this experimentally~? The first step
is to check the absence of fragmentation: 
A photo of the gas by absorption imaging should show, for any
realization of the experiment, that {\sl all} the particles are clustered in
a single lump of size $\xi$, randomly situated to the left or to the
right of the beam-splitter. The second step is to check that
the two wavepackets are coherent, by recombining them and looking
for interference fringes, with a fringe spacing $\pi/K_0$.
The recombination of the two wavepackets is obtained by their total
reflection on {\sl mirrors}, produced by two Gaussian laser beams centered
in $x=L/2$ and $x=-L/2$, $L\gg 1/\Delta K$,
with the same waist as the beam splitter but with a higher
intensity (say, twice as high). 
The reflected wavepackets interfere
around $x=0$, the beam splitter being switched off \cite{economy}.

\begin{figure}
\begin{center}
\includegraphics[width=8cm,clip]{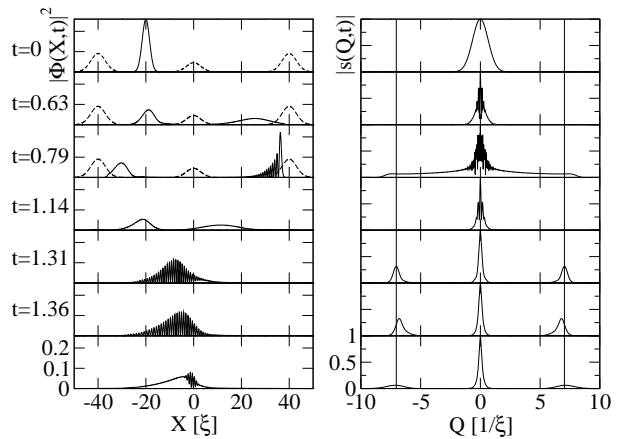}
\end{center}
\caption{
\label{fig:extrait}
Evolution of the center of mass wavefunction
of a solitonic gas with $N=99$ atoms, by integration
of (\ref{eq:se}) for the initial condition
(\ref{eq:init}), with
$\Delta K \simeq 0.093 K_0$, $X_0=-15\xi$.
Left panel: $|\Phi(X,t)|^2$ 
(solid lines), effective potential $\bar{V}(X,t)$ (dashed lines).
Right panel: modulus of the Fourier transform $s(Q,t)$
of $|\Phi(X,t)|^2$; the vertical lines are in $Q=\pm 2K_0$.
The time is in units of $ML/\hbar K_0$.
At $t=0.63$ the state is a cat.
At $t=1.31$ the two reflected cat components strongly interfere,
two narrow peaks of height $0.315$ emerge on $|s(Q)|$ in $Q=\pm 2K_0$.
Maximal interference occurs at $t=1.36$.
The last line is an average over a Poisson distribution
for $N$, with a mean value $\bar{N}=99$; 
the peak height in $|s(Q)|$ is reduced to $0.062$.}
\end{figure}

We have studied the proposed experiment
by a numerical solution of Schr\"odinger's equation for the center
of mass wavefunction, 
with the initial condition (\ref{eq:init}) and the same 
approximate effective Hamiltonian as in scattering theory:
\begin{equation}
i\hbar \partial_t \Phi(X,t) = \left[
-\frac{\hbar^2}{2M}\partial_X^2 + \bar{V}(X,t) \right] \Phi(X,t)
\label{eq:se}
\end{equation}
The center of mass
probability distribution $|\Phi(X,t)|^2$ is plotted at 
key times in Fig.\ref{fig:extrait}. 
To quantify
the contrast of the interference fringes, we also plotted
the modulus of its Fourier transform,
$s(Q,t) = \int_{-\infty}^{+\infty} dX\, e^{-iQX} |\Phi(X,t)|^2.$
When the two wavepackets overlap, sharp peaks in $|s(Q)|$
indeed form in $Q\simeq \pm 2 K_0$, with a contrast 
$|s|\simeq 0.32$.
This is a high value, as the ideal case of two overlapping plane waves 
$\Phi(X)\propto e^{iK_0X}+e^{-iK_0X}$ gives $1/2$.

The high contrast interference fringes in Fig.\ref{fig:extrait}
are however for the center of mass probability distribution, not for the atomic
density, which raises the question of their observability by usual
fluorescence imaging.
The mean atomic density $\rho(x)$ is the convolution of $|\Phi(X)|^2$ with
the internal soliton density $\rho(x|0)$; since the soliton size $\xi$
is as large as the fringe spacing $\pi/K_0$, one finds that the
contrast of the fringes in $\rho(x)$ is several orders of magnitude smaller
than in $|\Phi(X)|^2$. This problem can be solved by increasing, 
just before imaging,   
the intensity of the laser producing the transverse trapping
by a factor about 21, which reduces the transverse harmonic oscillator
length to $\tilde{a}_\perp=0.25\mu$m and brings the soliton close
to its collapse threshold $N|a|/\tilde{a}_\perp\simeq 0.67$
\cite{seuil}. Furthermore fluorescence imaging can be optimized 
to measure directly the quantity $|s(2K_0)|$,
by exciting the gas with a laser standing wave along $x$, produced
by the superposition of two laser waves of wavevectors $\vec{k}_\pm
=(\pm k_x, k_y,0)$
at some angle with the $x$ axis such that $\vec{k}_+-\vec{k}_-=2K_0 \vec{e}_x$.
The resulting fluorescence rate in direction $\vec{n}$ per unit
of solid angle is given in the Born approximation by
$d\Gamma/d\Omega\propto \langle 
\left|\sum_{i=1}^{N} e^{-ik\vec{n}\cdot\vec{r}_i} e(\vec{r}_i) \right|^2\rangle,$
with validity conditions discussed in \cite{Javanainen}.
Here $e(\vec{r}\,)$ is the laser electric field.
The emission rate $\Gamma_\Omega$ of photons 
in the solid angle $\Omega$ of the detection lens is an oscillating
function of the location of the antinodes of the laser standing wave
with respect to the interference pattern in $|\Phi(X)|^2$,
with a contrast 
\begin{equation}
\frac{\Gamma_\Omega^{\rm max}-\Gamma_\Omega^{\rm min}}
{\Gamma_\Omega^{\rm max}+\Gamma_\Omega^{\rm min}} =
|s(2K_0)| S_{\rm in}(\Omega).
\end{equation}
The reduction factor $S_{\rm in}(\Omega)$ is a function of the 
3D static structure factor of the soliton for fixed 
center of mass position, that we approximate with the 3D mean field
theory.
By using a lens of optical axis along $\vec{k}_++\vec{k}_-$
with a numerical aperture 0.4, one finds the remarkably
high value $S_{\rm in}(\Omega)=0.84$, thanks to a superradiant effect
\cite{Javanainen}, which also concentrates 16\% of the fluorescence
in the 4\% solid angle fraction collected by the lens.

It remains to check that decoherence is
negligible during the transit time $t_{\rm trans}=ML/\hbar K_0 
\lesssim 200$ms of the cat state in the interferometer.
In cold atom experiments,
the main source of decoherence is particle losses: 
A {\sl single} loss event would destroy the cat, since it ``measures" 
the positions of one or several atoms and localizes the center of mass
of the gas within the soliton size $\xi$.
The usual loss rate formula for $m$-body loss
is $dN/dt=-K_m \int d^3r\, n^m(\vec{r}\,)$; here one should take for
$n$ the 3D density profile for a fixed center of mass position,
that we approximate with the mean field theory.
For one-body losses due to
collisions with the background gas,
one should have a loss probability $K_1 N t_{\rm trans} < 1/10$, 
which imposes the reasonable lifetime $K_1^{-1}>200$s.
For three-body losses due to formation of deeply bound dimers,
the loss constant $K_3$ for $^7$Li at the considered magnetic field
$B$ is not known. Since $|a|$ is
smaller than the Van der Waals length $3$nm, as it is 
for $B=0$, we use the $B=0$ prediction of \cite{Verhaar}, applying
the factor $6$ reduction for a condensate,
$K_3 \approx 3\times 10^{-41} \mbox{m}^6/\mbox{s}$, which leads to
a negligible loss event probability $\frac{1}{3}|dN/dt|t_{\rm trans}
\approx 0.03$.

In present experiments the number of atoms $N$ fluctuates
from one realization to the other around the desired mean value $\bar{N}$.
Since the launch velocity $\hbar K_0/M$ is fixed, $K_0$ is proportional to $N$ and 
also fluctuates \cite{tbc}. A first side effect is that the half-transmission
probability condition may be violated away from $N=\bar{N}$;
fortunately this is not the case for a broad barrier $b\gg \xi$, since
both terms of (\ref{eq:htpc}) are proportional to $N$.
A second side effect is that the fringe spacing $\pi/K_0$ will fluctuate,
which will blur the fringes. A simple way  to estimate this
is to assume that $|\Phi(X)|^2\propto 
|e^{iK_0X}+e^{-iK_0X}|^2e^{-X^2/2\sigma^2}$ at the overlap time.
Averaging over a Poisson distribution for $N$ with 
$\sigma$ and $K_0/N$ fixed
leads to, for $|X|\ll \pi \bar{N}/\bar{K}_0$:
\[
\langle |\Phi(X)|^2\rangle \simeq \frac{e^{-X^2/2\sigma^2}}{(2\pi)^{1/2}\sigma}
\left[1+e^{-X^2/2\sigma_c^2}\cos(2\bar{K}_0X)\right].
\]
The fringes persist around the origin over a distance
$\sigma_c =$ $ \bar{N}^{1/2}/(2\bar{K}_0)= \sqrt{2}\xi.$
$|s(2\bar{K}_0)|$ is then reduced by a factor
$\sigma_c/(\sigma^2+\sigma_c^2)^{1/2}$. Estimating $\sigma$
from Fig.\ref{fig:extrait} leads to a reduction factor $5$
close to the numerical one
(see Fig.\ref{fig:extrait}).

In conclusion, we propose to produce a coherently bilocalized gas
by scattering an atomic quantum soliton on a barrier.
We have performed a detailed analysis of this idea, which raises challenging 
experimental aspects of preparation and detection, but also
non trivial theoretical aspects since this is a many-body problem.
We find that a gas with $N\simeq 100$ $^7$Li atoms 
can be prepared in a coherent superposition of being 
at two different locations separated by $\sim 100\mu$m, and that this
can be proved by an interferometric measurement.

One of us (C.W.) acknowledges financial support from the European Union
(contract MEIF-CT-2006-038407).
We acknowledge useful discussions with A. Sinatra, C. Salomon.
Our group is a member of IFRAF.



\begin{thebibliography}{99}
\bibitem{static}
This amounts to calculating for large $N$ the first
``beyond mean field" correction for the static structure factor 
$S(x,y|X)$ with {\sl fixed} center of mass position $X$.
The integral of $S(x,y|X)$ over $X$ is the structure factor
$S(x,y)$ of \cite{Caux} so our large $N$ expansion {\sl implies} 
the one of \cite{Caux}.
\bibitem{fluctu}
This splitting by filtering is more robust experimentally than macroscopic 
quantum tunneling: Fluctuations of the laser intensity producing the barrier
should only be such that resulting fluctuations of $K_0$ in (\ref{eq:htpc}) are 
$\ll \Delta K$.
\bibitem{economy} In a ring geometry,
recombination is provided for free by time evolution, and
our simulation results are similar.
\end{thebibliography}
\end{document}